\documentstyle[twoside,a4,12pt,epsfig]{article}
\def\be{\begin{equation}}
\def\ee{\end{equation}}
\def\bea{\begin{eqnarray}}
\def\eea{\end{eqnarray}}
\def\dps{\displaystyle}
\def\eeq{\endequation}
\def\beq{\beginequation}
\def\beq{\begin{equation}}
\def\eeq{\end{equation}}

\begin{document}

\title{SEARCH FOR PERIODIC GRAVITATIONAL WAVE SOURCES WITH THE EXPLORER DETECTOR}

%%%\author {{\rm ROG COLLABORATION}}
\author {
P. Astone$^1$, M. Bassan$^2$, P. Bonifazi$^3$, P. Carelli$^4$,\\
E. Coccia$^2$, C. Cosmelli$^5$, S. D'Antonio$^6$,
V. Fafone$^6$,\\ S. Frasca $^5$,
Y. Minenkov$^2$, I. Modena$^2$,\\
G. Modestino$^6$,
A. Moleti$^2$, G. V. Pallottino$^5$,\\ M.A. Papa $^7$, G. Pizzella$^{2,6}$,
L. Quintieri$^6$, F. Ronga$^6$, R. Terenzi$^3$, M. Visco$^3$ \\
{\it ${}^{1)}$ Istituto Nazionale di Fisica Nucleare INFN. Rome, Italy}\\
{\it ${}^{2)}$ University of Rome "Tor Vergata" and INFN. Rome, Italy}\\
{\it ${}^{3)}$ IFSI-CNR. Roma, Italy}\\
{\it ${}^{4)}$ University of L'Aquila}\\
{\it ${}^{5)}$ University of Rome "La Sapienza" and INFN. Rome, Italy}\\
{\it ${}^{6)}$ Laboratori Nazionali di Frascati-INFN. Frascati, Italy}\\
{\it ${}^{7)}$ Max Planck Institute for Gravitational Physics,AEI, Germany}
}

%\address{INFN, Rome ``La Sapienza", P. A. Moro 2, 00185
%\\ Rome, Italy }

\maketitle\abstract{}
We have developped a procedure for the search of periodic 
signals in the data of gravitational wave detectors.
We report here the analysis of one year of data 
from the resonant detector Explorer,
searching for pulsars located in the Galactic Center (GC).
No signals  with amplitude greater than $\overline{h}=
2.9~10^{-24}$, in the range 921.32-921.38 Hz, 
were observed using data collected over 
a time period of 95.7 days, for a source located
at $\alpha=17.70 \pm 0.01$ hours and $\delta=-29.00 \pm 0.05$ degrees.
Our procedure can be extended for 
any assumed position in the sky and for a more general all-sky search, even
with a frequency correction at the source due to the spin-down and Doppler
effects.

\section{Introduction}
Periodic or almost periodic gravitational waves (g.w.) are emitted by
various astrophysical sources.
They carry important information on their sources 
(e.g., spinning neutron stars, 
accreting neutron stars in binary systems) and
also on fundamental physics, since their nature can test the model of
General Relativity \cite{Schutz_sigrav,thorne}. 
The main feature of continuous signals which allows them to be detected 
is that, despite the weakness of the signal 
(compared to typical amplitudes for bursts),
it is possible to implement procedures that build up the
signal to noise ratio (SNR) in time. 
The natural strategy for searching for monochromatic waves 
is to look for the most significant peaks in the spectrum.
 In this case the SNR increases with the observation time $t_{obs}$. 
In fact, as $t_{obs}$ increases, the frequency resolution of the 
spectrum also increases - the frequency
bin gets smaller $\delta\nu=\dps{1\over t_{obs}}$ - thus the content
of the noise power in each bin decreases with $t_{obs}$, while the signal is
not dependent on the length of observation time\footnote{The noise
  power {\it{density}}, i.e. the power spectrum, is independent of
  $t_{obs}$, while the  
  signal energy {\it{density}} grows with $t_{obs}$}. 
For a periodic signal of amplitude $\bar{h}$ at the frequency
$\bar \nu$ the squared modulus of the Fourier Transform provides
${\bar{h}^2}$ with a noise contribution of $2~S_h(\bar{\nu})~\delta\nu$,
being $S_h(\bar{\nu})$ the two-sided noise power spectrum of the detector
%% (measured in $\dps {1\over Hz}$).
(measured in Hz$^{-1}$).

%The square modulus of the Fourier Transform \footnote{which is what
%we will refer to as the ``energy spectral density of the signal''} of
%a monochromatic signal at 
%frequency $\bar \nu$ and amplitude $\bar h$,
%has a pronounced peak at $\bar\nu$ and its height is $\dps{\bar
%  h^2\over 2\delta\nu}$. 

Thus the SNR
for periodic signals is:
\beq
SNR={{\overline{h}^2 t_{obs}} \over{2 S_h(\overline{\nu})}}
\label{snr_tuttonoto}
\eeq

Eq.\ref{snr_tuttonoto} holds if the 
instantaneous frequency of the continuous signal
at the detector is  known. The analysis procedure in this case is 
``coherent'', since the phase information contained in the data is used and
the sensitivity (in amplitude) increases with the square-root of the time.
However in some cases it may be impossible, for various reasons
(see later in sect. 3.1), to perform a single Fourier Transform over all
the data. This means that the observation time has to be divided in $M$
subperiods, such that the spectral resolution of the spectra becomes
$\delta \nu ^{'}=M / t_{obs}$ and the corresponding $SNR$ is $M$ times
smaller than that given by eq. \ref{snr_tuttonoto}.

The $M$ spectra can be combined together by incoherent summation, that is by
averaging the square modulus. In this case the final spectral resolution
is again $\delta\nu ^{'}$ but there is still some gain as this procedure
reduces the variance of the observed noise in the bin.
We obtain:
\beq
{SNR}^{\prime}={{\overline{h}^2 t_{obs}} \over{2 S_h(\overline{\nu}) \sqrt{M}}}
\label{snr_nontuttonoto}
\eeq

 In general, if the signal is
monochromatic but frequency modulated due to the
detector-source relative motion,  
processing techniques exist which can recover
the sinusoidal case if the source direction is known. One of the
standard ways of detecting such signals is through
appropriate  resampling of data, better known in the radio
astronomy comunity (where this technique is commonly used) as ``data
stretching'' (see for example \cite{camilo1}). 
In the case of radio pulsar searches, 
the location of the source is usually known (the data come from a
radio telescope pointing in a particular direction) 
but some parameters of the
system need to be estimated and this is done by
a ``timing solution
which is phase coherent over the whole data set''\cite{camilo1}. 

However for
gravitational waves, especially when searching a large parameter
space, it is doubtful that strategies developed for
radio pulsar searches can simply be adapted: the
expected low SNR values for g.w. signals really modifies the nature of the
search strategies that can be employed.
 In recent ms pulsar searches, for example in 
\cite{camilo1}, the signal is strong enough to allow suspected pulsars 
to be identified by {\it visual} inspection of the results of the final
stages of the analysis procedure. 
%power threshold at SNR=7 on data that had previously been
%de-despersed, FFTed and had undergone harmonic summing, the candidate 
%signals

%%The choice of the time length of the data to be used to compute the FFTs 
%%is a crucial step. 
%%Each of the FFTs in the data base has a sensitivity to the signal 
%%which is limited by its duration, say $t_0$. If
%%the analysis of the  FFTs is done  
%%``incoherently'', i.e. by averaging $M$ $|FFT|^2$, the final  sensitivity 
%%is given by \cite{pisa}

%%0\beq \overline{h}
%%=\sqrt {2S_h(\overline{\nu}) \cdot \delta{\nu}/ \sqrt{t_{obs} / t_0}}
%%\label{incoerente}\eeq
%%where $t_{obs}=M \cdot t_0$ is the total observation time,
%%$\delta{\nu}=1/t_0$ is the spectral resolution of
%%the FFTs.

For gravitational waves the study of the implementation 
of optimum analysis procedures is 
still in progress \cite{brady,map,krolak,aroma,brady2}.

In the present analysis,
we analyzed the data of the Explorer detector to search for 
ms pulsars located in the GC, 
assuming their intrinsic frequency to be constant over the analysis time.
%%a maximum time of $\simeq 10$ days.

The procedure we used in this study relies on a data 
base of FFTs, computed from short stretches of data 
(short with reference to the effects of the 
Doppler shift, as will be described later in this paper).
These short FFTs are then properly combined together 
to provide a new set of FFTs with higher frequency resolution, representing
the signal in the frequency range selected for the study.

The paper is organized as follows: in section 
\ref{expl} we briefly review the characteristics of the detector during the 
1991 run, in section \ref{main} 
we describe the main aspects of the procedure; in section
\ref{result} we present the results obtained.
%%Finally in section \ref{final} we discuss the extent to which the
%%constraints that we have introduced to describe the procedure can be
%%relaxed in order to account for different sources. We will also outline
%%future lines of investigation.
The Appendices clarify some aspects of the analysis procedure and
discuss the extent to which the
constraints that we have introduced to describe the procedure can be
relaxed in order to account for different sources.

\section{The Explorer detector}
\label{expl}
The Explorer detector is a cryogenic resonant g.w. antenna located
at CERN, at longitude $ 6^{\circ} {12}^{\prime}$ E and latitude
 $ 46^{\circ} {27}^{\prime}$ N.
The apparatus and the experimental set up of the antenna during the 1991
run have been described in \cite{longterm} and some results of the data
analysis for burst detection are given in \cite{phyrev}.

\begin{figure}
\centering\epsfig{file=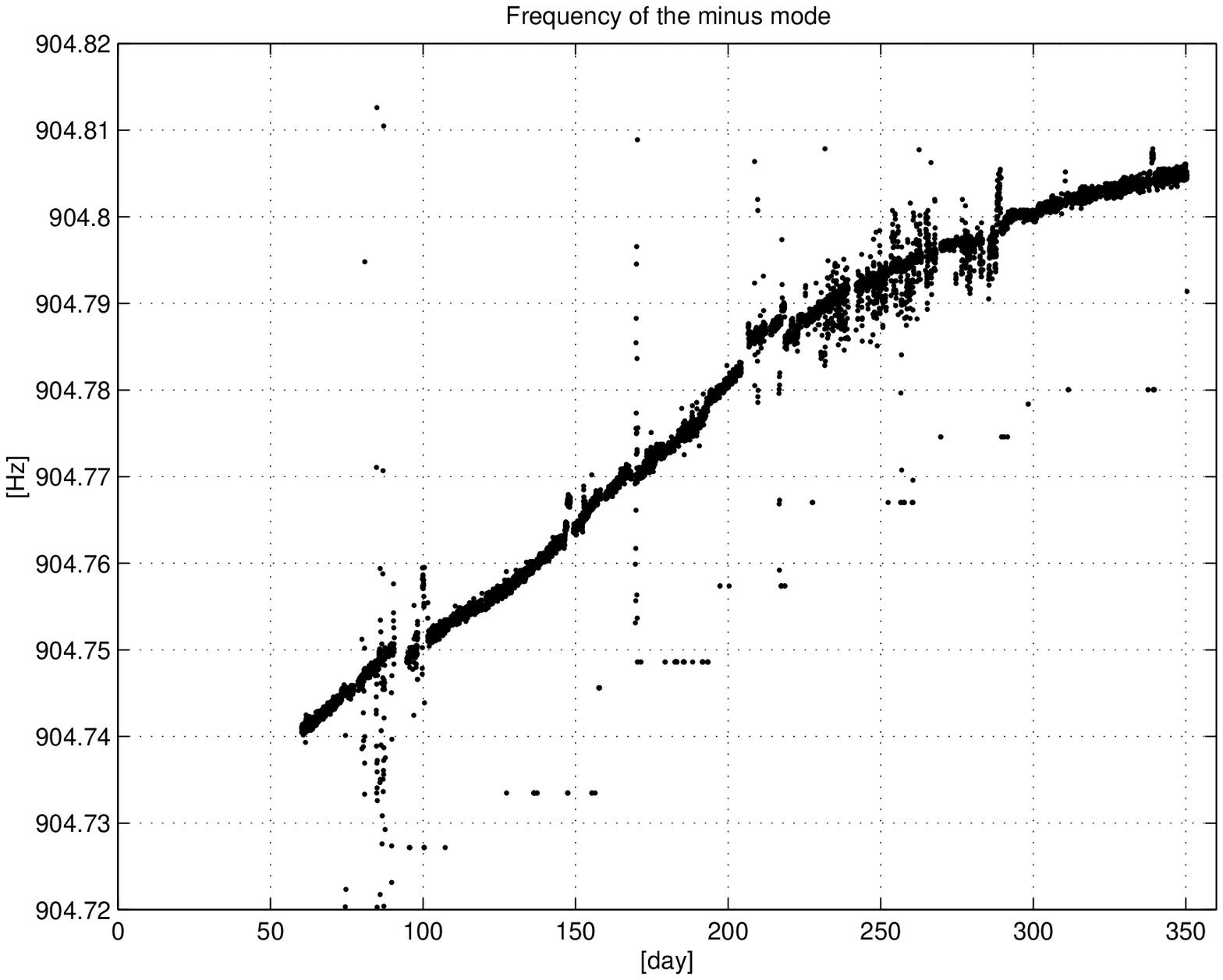,
width=8cm,clip=}
\centering\epsfig{file=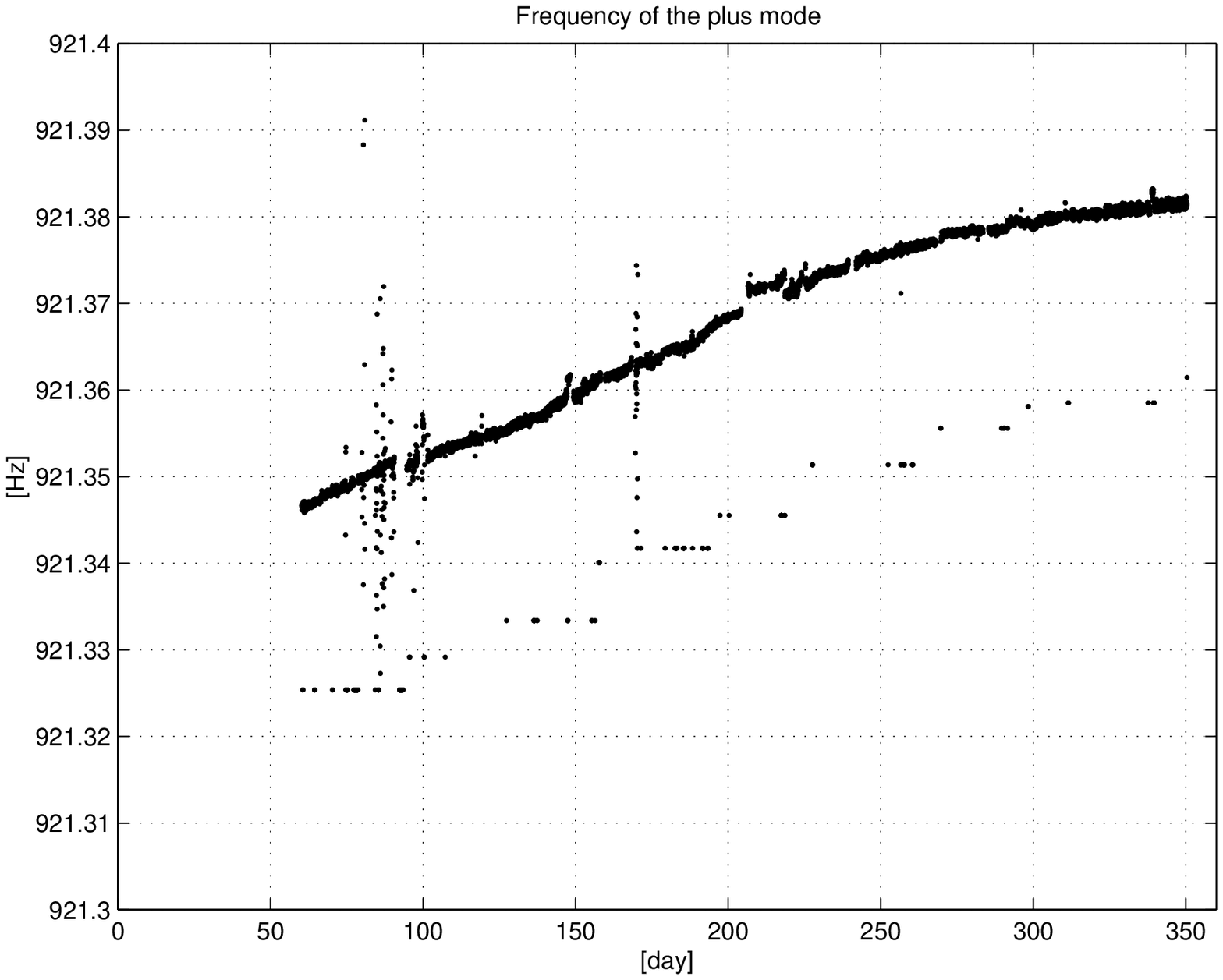,
width=8cm,clip=}
\caption{
Top: lower resonance (minus mode) frequency against time (in days of the year).
Bottom: upper resonance (plus mode) frequency against time.
A slow loss in the electrostatic charge of the transducer is probably
responsible for the frequency drift}
\label{fig:freqm}
\end{figure}

The system has two resonance frequencies
($\nu_-=904.7~Hz$ and $\nu_+=921.3~Hz$ in 1991) where the sensitivity is
highest.
Fig.\ref{fig:freqm} shows the behaviour of 
the two resonance frequencies with time. 
%Fig. \ref{fig:sigmam} shows the histogram of the brownian
%noise at the two resonances.
%The wide band noise has been measured over a bandwidth of 2 Hz between the
%resonances (1 Hz apart the calibration signal).

\par
\begin{figure}
\centering\epsfig{file=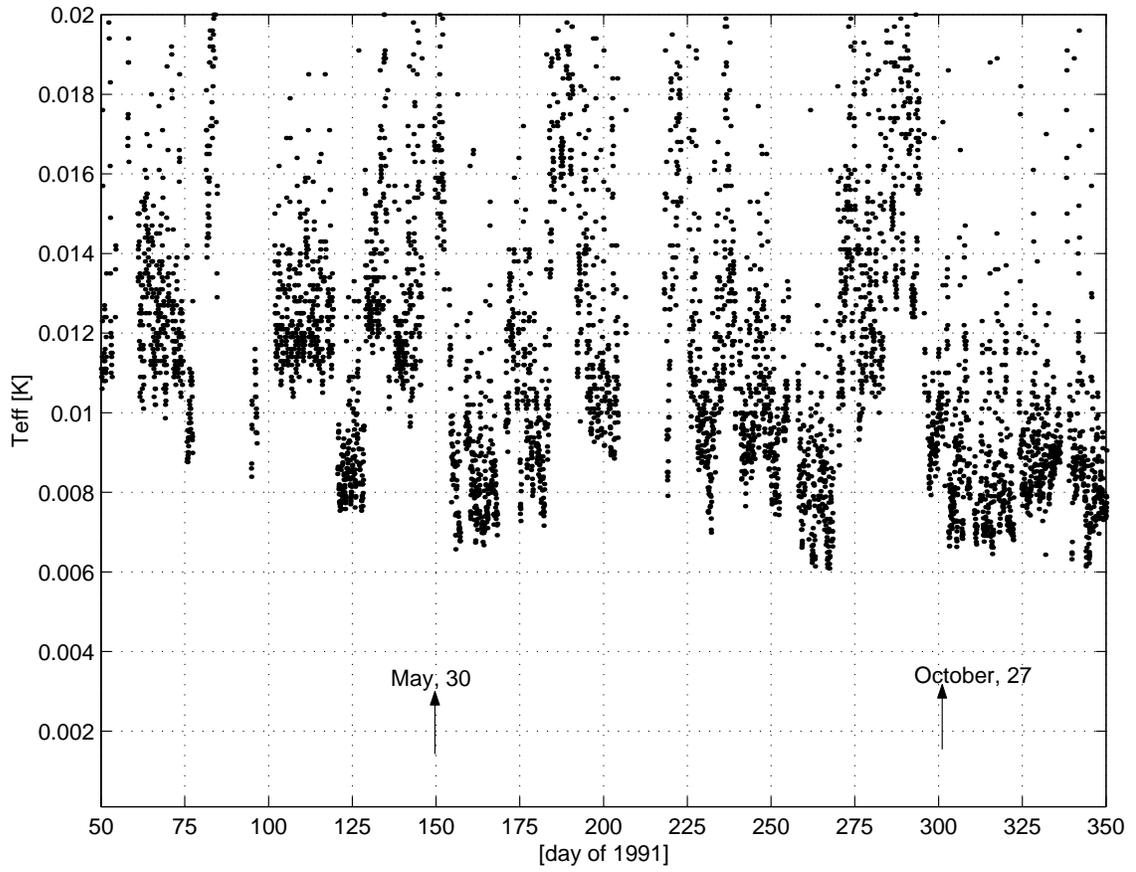,
width=15.0cm,clip=}
\caption{Hourly averages of the Explorer sensitivity to millisecond bursts,
expressed as noise temperature (K) 
as a function of time (in days of the year)}.
\label{fig:info91}
\end{figure}

Fig.\ref{fig:info91} shows the hourly averages of the 
energy sensitivity (SNR=1) to
millisecond bursts, expressed as effective temperature $T_{eff}$ in kelvin, 
obtained with an adaptive Wiener filter
\footnote{the sensitivity obtained with a matched filter was, on
average, better by a factor of two. The comparison between the
two filtering procedures is shown in \cite{veloce}.}.
The relation between $T_{eff}$ and the amplitude of a ms burst 
is \cite{longterm}
$h= 8~10^{-18} \sqrt{T_{eff}}$ ($T_{eff}$ in kelvin).

For periodic waves the sensitivity of a bar
detector at the above resonances 
is given by \cite{GVPGP,nostrocqg}:
\beq \overline{h}=2.04 \cdot 10^{-25} {\sqrt{ {T\over{0.05~K}} 
{2300~kg\over{M}}
{10^7\over{Q}} {900~Hz\over{\nu_0}} {1~day\over{t_{obs}}}
 }}\label{hcminimo}\eeq
where $T$ is the bar temperature, 
$M$ its mass, $Q$ the merit factor,
$\nu_0$ the resonance frequency of the mode and
$t_{obs}$ the observation time.
After one year of effective observation, 
the minimum detectable $\overline{h}$ (amplitude detectable with 
SNR=1), using the nominal parameters of the Explorer detector
($T=2$ K, $M=2300$ kg, $Q=10^6$), is
\beq 
\overline{h}=2 \cdot 10^{-25}
\label{idealexplorer}\eeq 
in a bandwidth of $\simeq 2~Hz$ around the two resonance frequencies
and $$ \overline{h} \simeq 2 \cdot 10^{-24}$$ 
in a bandwidth of 16 Hz between the two resonances. 
For the NAUTILUS 
\cite{nautilus} or AURIGA \cite{auriga} 
detectors (with $T=0.1~K,Q=10^7$) we get a value 
$\overline{h}\simeq 1.5 \cdot 10^{-26}$ at the resonances.

\section{Main features of the analysis procedure}
\label{main}

In the search for continuous signals there are a number of
  issues that need to be kept in mind regarding the signals that
  might be present, the apparatus and the quality of
  the data. As far as the source is concerned,
  it is not possible to
  set up a single procedure capable of searching over all types of
  periodic signals. In this analysis, we concentrated on  
  periodic signals such as those
  expected from isolated neutron stars with weak spin down, i.e. we ignored 
  the spin down parameters
\footnote{however it is possible, using our procedure, to take into
account the spin down. 
This will be the next step in the development of the procedure.}.
Moreover we did not consider
the effects of proper accelerations of the source.
Thus our model assumes that the frequency behaviour of the signal
exclusively depends on the Doppler effect caused by 
the motion of our Earth-based
detector relative to the source location. Let us quote some basic
figures:     

the Doppler effect has two periodic components 
(see Appendix A). 
The first one,
due to the revolution motion of the Earth over a  period of 1 year, produces 
a maximum spread of
\beq
\Delta \nu_{rev}=0.1986~ \overline{\nu} \cos{\beta_{ecl}}~~\mbox{mHz}
\eeq 
and a maximum time derivative
$
0.197 \cdot 10^{-10}~\overline{\nu}\cos{\beta_{ecl}}~~
\mbox{Hz} \, s^{-1}
$
where $\overline{\nu}$ is the intrinsic frequency of the signal and 
 $\beta_{ecl}$ is the ecliptical latitude of the source. 
The second one, due
to the rotation of the Earth over a period of 1 sidereal day, produces 
a maximum spread of
\beq
\Delta \nu_{rot}=0.00308~\overline{\nu} \cos{\beta_{ter}} \cos{\delta}~~
\mbox{mHz}
\eeq 
and a maximum time derivative
$
1.12 \cdot 10^{-10}~\overline{\nu}\cos{\beta_{ter}}
\cos{\delta}~~\mbox{Hz} \, s^{-1}.
$
$\beta_{ter}$ is the terrestrial latitude of the detector and $\delta$
the declination of the source. 

The observation is also affected by
modulation in the amplitude.
 This is due to the varying orientation of the
detector with respect to the source because of the Earth's motion. It may also
be a consequence of the polarization of the wave.
 
As shown, for example, in \cite{livas},
this modulation spreads the signal power across side bands, spaced
at 1/24 hours. 
%mostly at about
%$\pm 0.116$  mHz and $\pm 0.232 $ mHz. Typically these sidebands contain
%$1/3$ of the total radiated power. 
The amplitude modulation observed using a resonant bar detector is given by
the geometrical part of the detector cross section \cite{sergio}
\beq
\Sigma=\Sigma_0 \cdot \Phi (\theta,\epsilon,\phi)= \Sigma_0 \cdot
sin^4 \theta \cdot ({{1-\epsilon} \over{2}} + \epsilon cos^2(2 \phi))
\label{sezioneurto}\eeq
where $\epsilon$ is the percentage of the wave linear polarization 
($\epsilon$=1 when there is linear polarization,
$\epsilon$=0 when there is no polarization),
$$\Sigma_0={16 \over {\pi}} ({{v \over {c}})^2} {G \over {c}} M~~~~[m^2~Hz]$$
is the (two-sided) mechanical part of the cross section,
($M$=bar mass, $v$= sound velocity in the bar),
$\theta$ is the angle between the bar axis and the wave 
direction of propagation,
$\phi$ is the angle between the bar axis and the wave 
polarization plane.
The cross section is maximum when the parameters are
$\theta=\pi/2, \phi=0~, \epsilon=1$.
If the source location and the polarization state are known, 
it is possible to demodulate the amplitude of the
observed signal.

\begin{figure}
\centering\epsfig{file=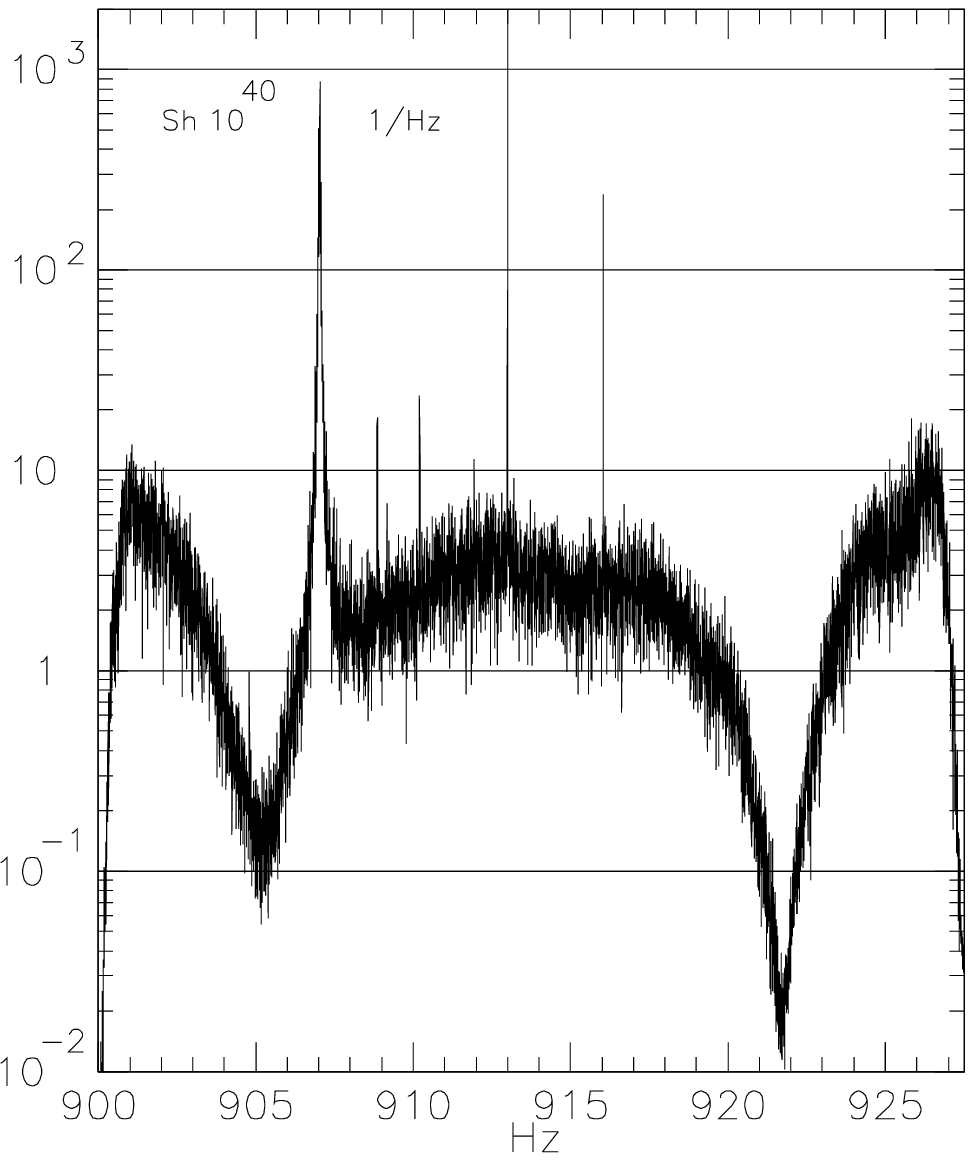,
width=5.0cm,clip=}
\centering\epsfig{file=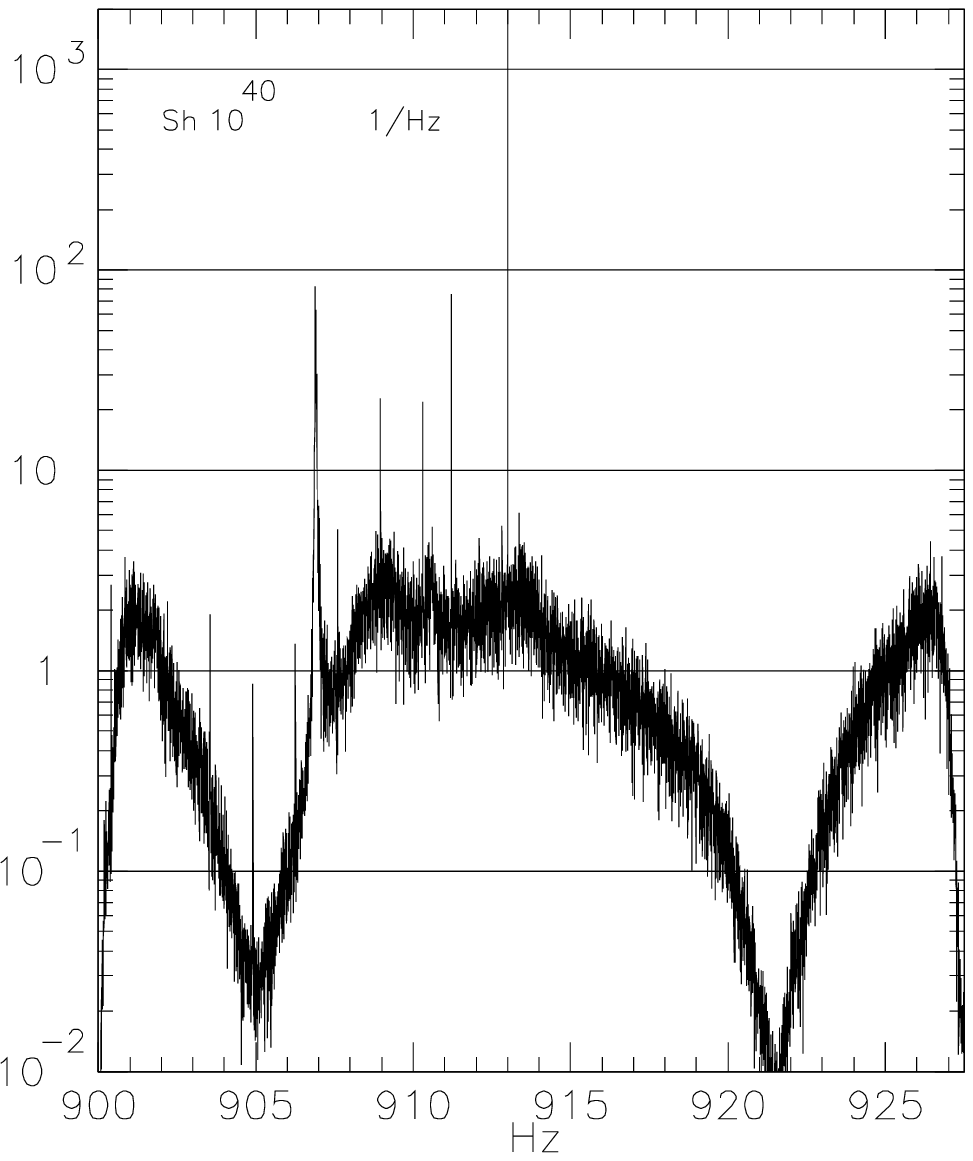,
width=5.0cm,clip=}
\caption {The figure shows two power spectra of the detector, both obtained
during periods of ``good'' operation of the apparatus:
during November (left) and September (right) 1991.
The y-axis is $S_h \cdot 10^{40}$ in units of $1/Hz.$
The x-axis is the frequency in [Hz]. 
Comparing the two power spectra, it is easy to see 
the non stationarity of the system}
\label{fig:smoothingaa}
\end{figure}

A major consideration in developing the analysis is that the
operation of the detector is not continuous and the noise is not
stationary. 
An example of this is given 
in Fig.\ref{fig:smoothingaa} which shows power
spectra computed over two hours of data 
in November and September 1991. There are several
lines from periodic disturbances which are not stationary, and the noise level
differs between the two spectra.

A final remark needs to be made about the time accuracy required 
to perform a coherent search over a given time period.
A rough calculation, shown in Appendix B, shows that, in 
a coherent search, the global timing error 
must be less than $10^{-4}-10^{-5}~\mbox{s/year}$.
%%%%%%%****ovviamente per fare l' analisi su un anno

\subsection{The length of the FFTs in the data base}
Our frequency domain data base consists of ``elementary spectra'', each
obtained by performing the FFT (Fast Fourier Transform) of a given number of
samples of the data, over a duration $t_0$, recorded by our detector. The
signal sensitivity of each spectrum, according to eq. \ref{snr_tuttonoto},
depends on the duration $t_0$.
As our observations are affected by the Doppler shift, we have chosen the
duration $t_0$ of the ``elementary spectra'' to be
the longest possible compatible with the
requirement that the signal should "appear as monochromatic"
during $t_0$. This is expressed
as:
\beq
t_{{\rm{0~opt}}}={\rm{max~}}t_{\rm{0}} {\rm ~~such~that~~} 
\int_{t_{\rm{0}}}\dot{\nu}(t) ~dt \le
\dps{1\over 2 t_{\rm{0}}}.
\eeq
Clearly some assumptions about the expected $\dot{\nu}$ must be made. In
principle, in order to achieve a higher SNR in
the short spectrum, some preprocessing could take place by setting a
coarse grid on the parameter space 
(i.e., the part of the sky that is being investigated)
and performing suitable data stretching
for each point in that parameter space.
In this way a signal coming from that
parameter space would appear as monochromatic in the
resulting spectrum. 
As a consequence the dimension of the data base is increased
by a factor equal to the number of points in parameter space, but there
is a gain
in SNR because of the higher spectral resolution. 
%As pointed out by
%P. Brady et al. (\cite{brady,brady2}), there is an optimum length and an
%optimum number of spectra that can be analyzed incoherently,
% for any given computational resource.

As stated above, we restricted our analysis to the case where the
only frequency changes are due to the Doppler effect of the detector
motion relative to the source. 
%We have not
%considered, for the time being, more precise phase corrections which would
%take into account the area in the sky where any particular source is
%located.

%%The incoherent overall sky search analysis of the short spectra 
%%is still in progress.
As shown in Appendix A,  
the time duration must be
\beq
t_0 <\frac{1}{\sqrt{a+b}} \,
= 8.6972 \cdot 10^{4}/\sqrt{\overline{\nu}~}~~s \, , \label{dausare}
\eeq
with
%where
%$1/T$ is the spectral resolution in each FFT computed from stretches
%of data of length $T$,

\noindent
$a= \overline{\nu} \cdot 11.244 \cdot 10^{-11} \cos \beta_{ter}  ~
\mbox{Hz} \, s^{-1}$ 
(daily rotation),

\noindent
$b= \overline{\nu} \cdot 1.977 \cdot 10^{-11} \cos \beta_{ecl} ~
\mbox{Hz} \, s^{-1}$ 
(annual orbital motion),

\noindent
and $\overline{\nu}$ is the source intrinsic frequency.

%At 1 kHz, over a time $t_0\simeq 40$ minutes, T

Thus, to construct the elementary spectra of our data base, we chose 
a duration of $t_0=2382.4$ s=39.7 minutes, corresponding to
$2N=131072$ samples, recorded with sampling time of $18.176$ ms.
With this choice of $t_0$, the maximum Doppler frequency
variation, $\delta{\nu_{d}}=0.28$ mHz, is smaller than the
resulting frequency bin $\delta{\nu}=0.419$ mHz.

%%The Explorer data were sampled in
%%a bandwidth of $27.5$ Hz in the frequency range
%% $900$ - $927.5$ Hz, with sampling time  $ 18.18...$ ms. Thus
%%in the time $t_0=0.66$ hours each FFT has $2N=131072$ samples.

%For a resonant detector, operating roughly at 1 kHz, due to the
%Doppler effect 
%(considering only the Earth motion) the maximum
%sweeping in frequency is $\simeq 0.62$ Hz during one year and
%$\simeq 0.011$ Hz during one day. 

\subsection{The FFT header}
\label{header}

The header of each elementary spectrum of the database contains various
information about the original data.
This allows stretches of data that are noisier than others 
to be vetoed or weighted differently and thus best exploits the 
potential of the data.
  
Some of the information contained in the header 
relates to the data structure, some of it to the operational status of
%%%data structure puo' essere detto meglio ?%%%%%%%%
the detector and some to data quality. For example,
the date and time of the first sample of the data series that the
FFT is computed from is stored, along with the frequency resolution
of the FFT and the type of time-domain windowing used. There
are also system parameters that vary in
time: the frequencies of the two modes and of the calibration signal,
the level of
brownian noise and merit factors at the two modes, the 
wide band noise level and the 
status of the operation flags (normal operation, working around the antenna,
liquid helium refilling). 
Some of this information was used to set a threshold for vetoing data
(at  40 minutes, which is very reasonable due to typical
non stationarities of the detector)

\subsection{The procedure for combining spectra}
\label{procedure}

For the targeted search described here the basic FFTs are combined 
coherently to improve the final sensitivity. 
The following is an outline of how this is done 
(details are given in Appendix C).

\begin {itemize}

\item
take an FFT.
%, be this the $i^{th}$ FFT of the chosen set 
Let $\delta\nu$ be the frequency resolution, $2N$ the number of samples and $B$ 
the bandwidth of the detector;

\item

take the data from $n'$ bins in the frequency band  $\Delta \nu$ of the
actual search; $n'= N
\Delta \nu / B;$

\item
build a complex vector that has the following structure: 

- the first datum equal to zero 

- the next $n'$ data equal to those from the selected bins of the FFT

- zeroes from bins $n'+1$ up to the nearest subsequent bin 
numbered with a power of two. Let us say that this
way we have $n$ bins.

- zeroes in the next n bins

So, we end up with a vector that is $2 n$ long.

\item 
take the inverse FFT of the vector. This is a complex time series
that is the ``analytical signal'' representation of the signal in
the band $\Delta \nu$.
 It is shifted towards lower frequencies and it is sampled at a
sampling rate lower by a factor $2N/2n$ compared to the original time data
\footnote{
the construction of the analytical signal is a standard
procedure of lowpass filtering for a bandpass process. In fact the
analytic signal is zero on the left frequency plane, thus avoiding
aliasing effects in the lowpass sampling operation \cite{tretter}}.
The time of the first sample here is exactly the same as the first
datum used for  the data base and the total duration is also that of
the original time stretch. There are fewer data because the sampling
time is longer.

\item
repeat the steps outlined above for all the $M$ FFTs; 

\item
if they all come from contiguous time stretches simply append one 
after the other in chronological order. If they are not all contiguous 
set to zero those stretches where data are missing.

\item 
To correct for the Doppler effect 
\footnote{
we could also take into account other causes of frequency
shifts such as those affecting the intrinsic
frequency of the source.} 
from a given source, multiply each
sample of the sequence by 

\beq
exp^{ (-j \phi(t_i))}
\label{correctingfactor}
\eeq
$t_i$ are the times of the samples and
$\phi(t_i)=\int_{t_s}^{t_i} {\Delta \omega_D(t) dt}$.

$\Delta \omega_D (t)$ is the Doppler correction, in angular
frequency, at the time $t$ of the $i^{th}$ sample:
$\Delta \omega_D (t)=\omega_D(t)-\omega_s$, where $\omega_D(t)$
is the frequency observed at the detector, due to the Doppler effect
from a given source that emits at a constant frequency $\omega_s$.
$t_s$ is the start time of the FFT being constructed.

We note that the frequency correction is performed on
the subsampled data set, and this is one of the advantages of the procedure.

\item
Perform the FFT of the $2n \cdot M$ data thus obtained.

\item 
Finally, take the squared modulus of the FFT thus
obtained. This is the power spectrum of the original time series, in
the band $\Delta\nu$, with the full spectral resolution $\delta \nu /
M$. A signal exhibiting frequency variability smaller
than the variability we have corrected for, should appear 
wholly within a single frequency bin, and its resulting SNR will  be that 
of Eq.\ref{snr_tuttonoto}

\end{itemize}

\subsubsection{An example of the procedure for combining spectra}
The procedure was tested on simulated signals added to the
data. We shall now briefly review the results of these tests.
Such simulations, although simple 
in principle, present 
practical design problems which demand extreme care in the implementation.
 The simulated signal is constructed in the time domain and then it is 
handled in exactly the same way as the real detector data
(details on the data handling procedures are given in the Appendices). 
Each FFT of the signal is then added to the
corresponding FFT in the data base.
\beq
s(n \Delta t)=\overline{h}(n \Delta t) \, \sin{(\phi(n \Delta t)+ \phi_0)}
\eeq
where 

\noindent
$\Delta t$ is the sampling time, $n=0,1....131072$, $\phi_0=$initial phase,

\beq
\phi(n \Delta t)=\int_{0}^{n \Delta t} {\omega_D(t) dt}
\label{inifase}
\eeq

\noindent
where $\omega_D(t)$ is the frequency at the detector due to the Doppler
shift at time t.
Using the discrete form of eq.\ref{inifase} we may write
the phase at time $t_i$:
$$\phi_i=\phi_{i-1}+\omega_{Di} \, \Delta t$$

We report here an example in the absence of noise 
(we set the detector FFTs to zero, before adding them to the simulated signal).

\begin{figure}
\centering\epsfig{file=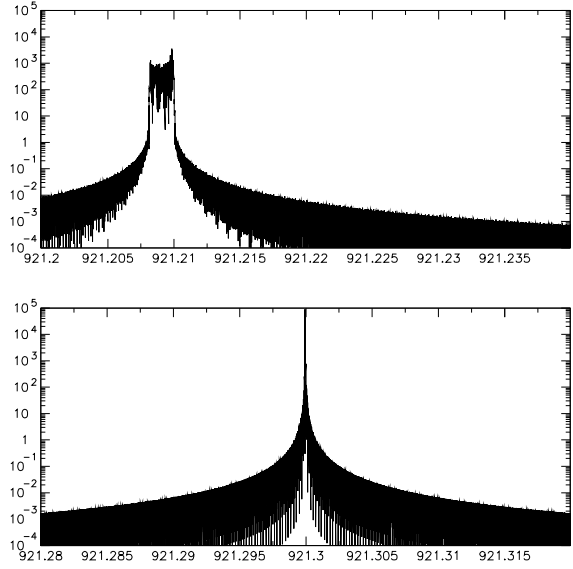,
width=8.0cm,clip=}
\caption{Simulation of a signal (at 921.3 Hz) from the GC, 
over 36 hours of data. Top: $S_h(\nu)$ of the simulated data.
Bottom: $S_h(\nu)$ after Doppler removal.
The y-axis is $S_h \cdot 10^{40}$ in units of $1/Hz.$
The x-axis is the frequency in [Hz]. }
\label{fig:sim}
\end{figure}

Fig.\ref{fig:sim} shows 
the comparison of the two power spectra obtained
from a source assumed to be in the GC, emitting at 921.3 Hz, before 
and after Doppler removal. It is clear that the spread and the
shift in the signal frequency (top figure) have been 
properly corrected (bottom figure). 
Here the observation time is 36 hours and the frequency resolution 
is $6.4... \mu Hz$. 

The level of the signal, after Doppler removal, is that which would be
expected  ($\overline{h}=1.0 \, /\sqrt{Hz}).$
An accurate analysis of the residual error after Doppler removal
(this error is defined 
as the instantaneous difference between the time signal after
correction and the time signal in the absence of modulation) showed
that this residual error was always less than 
 $0.7 \%$. 

\par
\newpage
\section{The analysis of the Explorer data}
\label{result}

We analyzed data taken during the period between March and December 1991. 
These data sets comprise 4954 FFTs from March  to July and 
4384 from August to December.
After a preliminary analysis of the features of the spectra, with particular
reference to their sensitivity performance, we
decided to veto the spectra with brownian noise larger than
7.8 K (i.e. three times greater than the
expected value of 2.6 K \cite{longterm}),
With this criterium we vetoed 807 spectra, that is $\simeq 10$ \% of the total.

%\begin{figure}
%\centering\epsfig{file=sigmam_tutto.eps,
%width=8.0cm,clip=}
%\centering\epsfig{file=sigmap_tutto.eps,
%width=8.0cm,clip=}
%\caption{Histograms of the
%Brownian noise at the two resonance modes (top:minus, bottom:plus) expressed as
%$\phi_0^2~10^9$. The peak at low values, in both cases, is due to the
%runs in May and in June, when the detector sensitivity
%was at its best}
%\label{fig:sigmam}
%\end{figure}

%\par
%\begin{figure}
%\centering\epsfig{file=sigmaw.eps,
%width=6.0cm,clip=}
%\centering\epsfig{file=sigmaw2.eps,
%width=6.0cm,clip=}
%\caption{left: March-July 1991, right:August-December 1991.
%The top figures show the wide band noise, expressed as
%$\phi_0^2/Hz \cdot 10^12$. The bottom figures show their distributions}
%\label{fig:sigmaw}
%\end{figure}

%Analyzing of the histograms, we decided to veto all the FFTs with

%$\sigma_m^2 > 60 \cdot 10^{-9} \phi_0^2$,
%$\sigma_p^2 > 40 \cdot 10^{-9} \phi_0^2$ 
%%%$\sigma_w^2 > 1.2 \cdot 10^{-9} \phi_0^2$.

A comment on the accuracy of the timing of the data
is necessary at this point:
the absolute time recording had an indetermination
of the order of $10-20~ms$ \cite{longterm} 
at the beginning of each new run. 
This was due to the fact that, 
although the time was checked against the Swiss time signal HBG 
with an accuracy of a few ms,  
the software procedure at the start of each run introduced an imprecision of 
$\simeq 10-20$ ms.

On the other hand  we are  confident about the precision of the 
Rubidium clock, which was used to determine  
the sampling time. 
%%%%%%%%%(and had a stability
%%%%%%%better than **********/day ----numero da chiedere !!!). 
As a consequence we could combine 
coherently only data obtained from one single acquisition run.

The strategy for the analysis procedure was thus the following:

\begin{itemize}
\item
choose the frequency bandwidths to be analyzed,
and calculate new -higher resolution- 
FFTs on each new run, for each of these chosen bandwidths
\item
choose the coordinates of the source location and correct for Doppler
 effect and for amplitude modulation, 
using the procedure described in section \ref{main}. 
 We ignored possible polarization of the waves.

\end{itemize}
This  analysis was focused on possible sources in the GC,
at $\alpha=17.7$ hours, $\delta=-29.0$ degrees.

To calculate the Doppler shift we used the JPL
ephemerides (JPLEPH.405) and software routines from the
US Naval Observatory (NOVAS).
The amplitude modulation was removed from the data by 
matched filtering, multiplying the data by the 
factor $\sin^4 \theta (t)$ (in other words, the data were weighted
on the basis of the source-detector direction).

In 1991 we collected data over 51 separate runs and therefore,
applying the procedure outlined above, we obtained 51 separate FFTs. 
Each one has a different frequency resolution, according to its length in
time (due to the timing error between different runs).
The analysis of the 51 FFTs could only
 be done by combining their information
``incoherently'', thus combining the spectra. 
Obviously, this reduced the sensitivity of the final analysis
(see Eq. \ref{snr_nontuttonoto}). 
 
We report here the results of the analysis of the data around the 
frequency of the plus mode
%%, for the 
%%bandwidth $921.20-921.45$ Hz 
(the mode in Fig.\ref{fig:freqm}, bottom).
%%\footnote{In practice we had to choose a 
%%wider bandwidth, to avoid problems when considering the Doppler corrections
%%at the edges of the selected data. Also we needed
%%to have a number of data equal to a power of two,
%%to minimize the analysis time.}

\subsection{The analysis of 95.7 days of data}

First of all we give an example using data over 
one week in June 1991.
%Fig.\ref{fig:figure302z} shows one spectrum, obtained over 
The observation time is
$t_{obs}=7.05$ days from day 159.8 (June, 8th). The analyzed data are
in the bandwidth 921.32-921.38 Hz.
The level of the noise is $(1.2 \pm 0.7)~ 10^{-24}$,
%, obtained from the noise spectral density using 
% Eq.\ref{snr_tuttonoto},
in good  agreement with that expected for Explorer 
%%at a temperature of 2.6 K 
(using eq.\ref{snr_tuttonoto}, with $T=2.6$ K and $Q=10^6$, 
we get $1.9 \cdot 10^{-24}$)
%%, which is smaller
%%than the average Explorer temperature of 2.6 K but within the observed
%%fluctuations. Indeed during the
%%7.07 day period the measured noise thermal temperature was below 1 K.

The Doppler correction needed for signals from the GC 
was applied to the data. The highest peak is 
$\overline{h}=5.2~10^{-24}$. %%%%%%%%at the frequency $\nu=921.354...$ Hz.

%\par
%\begin{figure}
%\centering\epsfig{file=figure302.eps,
%width=10cm,clip=}
%\caption{One spectrum obtained
%over 7.05 days (from june, 8th), in the
%bandwidth 921.2-921.45 Hz. The y-axis is normalized to give the
%amplitude of signals detectable with SNR=1.
%The Doppler correction from GC has been applied on the data.} 
%\label{fig:figure302}
%\end{figure}

%\begin{figure}
%\centering\epsfig{file=figure5_new.eps,
%width=10cm,clip=}
%\caption{Amplitude ${\overline{h}}$ of signals detectable with SNR=1,
%over 7.05 days (from June, 8th), in the
%bandwidth 921.32-921.38 Hz.
%The data have been corrected to look for signals from the GC. 
%The x-axis is the frequency in units of Hz, having subtracted 921 Hz.}
%\label{fig:figure302z}
%\end{figure}

%%For comparison we show in Fig.\ref{fig:figure303z} the case where
In the case that
no Doppler correction was applied to the data there is
a very high peak, at the level $\overline{h}=1.2~10^{-23}$,
which disappears when the Doppler correction based on
a source in the GC is applied. It is most likely that the
peak 
%%%in Fig.\ref{fig:figure303z} 
was due to the apparatus. 
The GC Doppler correction spreads this contribution over several
frequency bins.

%%\begin{figure}
%%\centering\epsfig{file=figure6_new.eps,
%%width=10cm,clip=}
%%\caption{The same data as for Fig.\ref{fig:figure302z}
%% but in this case no Doppler correction is applied. 
%%The high peak disappears when applying the Doppler correction, 
%%thus it is not due to monochromatic signals from the GC.} 
%%\label{fig:figure303z}
%%\end{figure}

%%Since the final sensitivity is
%%both a function of the observation time and of the noise level,
We started the analysis using only 
the data from three consecutive runs in May and June.
%% when the measured noise temperature was at a good level, compared to
%%the average noise of the whole period.

%%%%%%%%%Figure \ref{fig:figure304} shows the spectrum 

\begin{figure}
\centering\epsfig{file=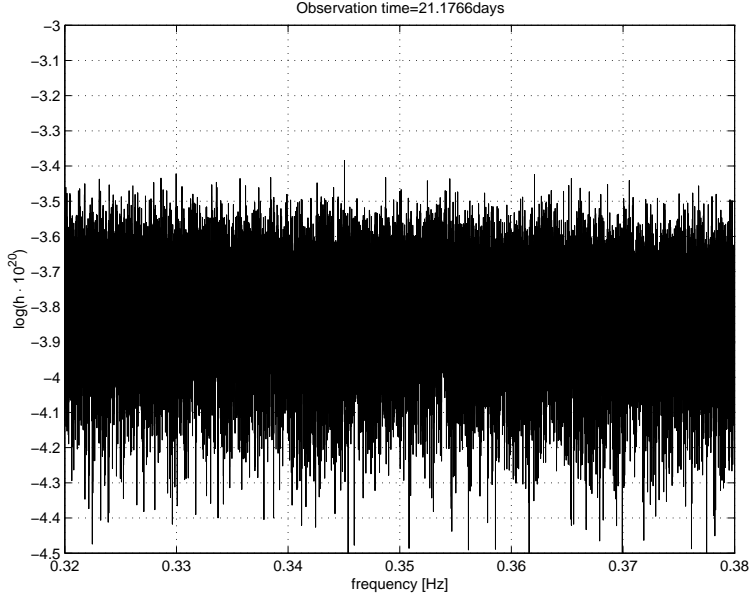,
width=10cm,clip=}
\caption{Average $\overline{h}$  from GC obtained averaging the spectra 
of 3 runs, 7 days each, from May, 8th. 
The x-axis is the frequency, in Hz-921 Hz.} 
\label{fig:figure304z}
\end{figure}

We averaged the corresponding spectra over a total observation time of 
$t_{obs}=21.177$ days from day 128.53 (May, 8th). The  
analysis was done in the frequency 
range 921.32-921.38 Hz,
where the antenna noise spectral density was flat, as shown in 
Fig.\ref{fig:figure304z}.
The figure  is normalized in terms of the amplitude $\overline{h}$
that would give $SNR=1$ for sources in the GC.
The  $\overline{h}$ level is $(1.6 \pm  0.5)~ 10^{-24}$. 
We notice that the average level
for the 21.177-day period 
%%(see fig.\ref{fig:figure304z})
is roughly (within the observed non stationarities) 
the same as that for the 7.05-day period,
and the standard deviation decreases, as it should do, by a factor
of the order of $\sqrt{\frac{21.177}{7.05}}=\sqrt{3}$
(the small difference is due to the non stationarities during the 
three time periods).
\par
%% e' il complessivo e qui non serve
%%\begin{figure}
%%\centering\epsfig{file=figure304.eps,
%%width=10cm,clip=}
%%\caption{Average spectrum obtained from data in may and june 1991
%%over 21.177 days, in the
%%bandwidth 921.2-921.45 Hz. The figure is normalized to give the signal
%%amplitude and the Doppler correction from GC has been applied on the data.} 
%%\label{fig:figure304}
%%\end{figure}

No spectral lines were detected with amplitude (at the detector) greater
than $\overline{h}=4.1~10^{-24}$. %%%$\nu=921.345...$Hz.

To set an upper limit on the amplitude of possible signals from 
GC in the chosen bandwidth, we decided to check 
the efficiency of detection, given the noise of the detector.
We therefore added signals 
with different amplitudes and phases to the data, 
using data without Doppler correction, since the
efficiency of detection, on the average, does not depend on
the Doppler effect.

%%esattamente sono 1.4 e 2.6.
We added four different families of signals, each family consisting of
20 sinusoids with the same amplitude but different phases.
%%%%%%%%%%% 921.34999 Hz

\begin{figure}
\centering\epsfig{file=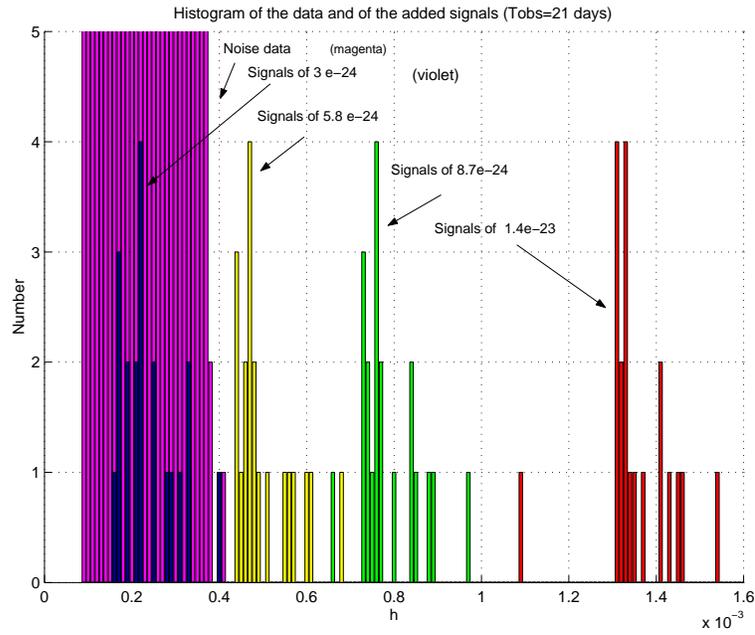,
width=10cm,clip=}
\caption{Histograms in the bandwidth 921.32-921.38 Hz.
$t_{obs}=21$ days.
The histograms report the data (magenta) and the four different families
of simulated signals (violet,yellow,green,red). 
The y-axis numbers above five have not been plotted. 
The x-axis ranges from
$1.0~10^{-24}$ to $1.6~10^{-23}$, with interval $0.1~10^{-24}$. 
The histograms show clearly that the efficiency of detection for
 the simulated signals corresponding to the yellow, green and red plots is 1.} 
\label{fig:isto}
\end{figure}

Fig.\ref{fig:isto} shows the histograms.
%% for the frequency range 921.32-921.38 Hz.
The histograms report the data and the four different families
of simulated signals. On the
x-axis we have values ranging from $\overline{h}=
1.0~10^{-24}$ to $1.6~10^{-23}$, with an interval of  $0.1~10^{-24}$.

The nominal amplitudes of the added signals are 
$1.4 \cdot 10^{-23}$, $8.7 \cdot 10^{-24}$, $5.8 \cdot 10^{-24}$
and $3.0 \cdot 10^{-24}$.
If we consider only the three families of higher signals,
the histograms show very clearly that all these 
signals have been well detected. 
Even the smallest of these 60 signals is well above 
the standard deviation of the data 
(the smallest three signals  
have $\overline{h}= 4.4~10^{-24}$, which is 
roughly 10 times the  noise standard deviation).
Thus, for these signals, on a time basis of only 21 days, 
the efficiency of detection is 1.
On the contrary, the histograms show that
the efficiency of detection for the signals at the lowest level 
($3.0 \cdot 10^{-24}$) is very poor.
%To have an idea of the effect of the noise on these signals we report here
%their average values and the standard deviations, that are, respectively:
%%%att:cambiati senza lo smoothing%%%%%%
%$\overline{h}=4.7 \pm 0.3 \cdot 10^{-24}$;

%$\overline{h}=2.7 \pm 0.2 \cdot 10^{-24}$;

%$\overline{h}=1.7 \pm 0.2 \cdot 10^{-24}$.

Thus, on the basis of 21 days of data, 
we exclude the possibility that, in the GC, 
there are sources 
having a spindown age $\tau \ge 3 \cdot 10{^7}$ years
%%(that is $\dot{p} \le 5 \cdot 10^{-19}$ s/s).  
%questo dovrebbe essere 21 giorni icoerenti mediando 3 volte 7 gg
emitting signals with frequency 
in the range 921.32-921.38 Hz and strength (on Earth) greater than or equal to
$\overline{h}=5.8~10^{-24}$.

%%However we have to consider the possibility that the used periods of small
%%thermal noise were really affected by a systematic error, bringing
%%the measured temperature from the average value of 2.6 K to the smaller
%%value of 0.8 K. If we apply a correction then the value
%% $\overline{h}=2~10^{-24}$ becomes $\overline{h}=
%%\sqrt{\frac{2.6}{0.8}~2~10^{-24}}=3.6~10^{-24}$.

For the eleven longest runs between May and December 
the observation times range from 7.7 to 12.8
days, giving a  total effective observation time of 95.72 days.
We can average these spectra, after adding 
the necessary zeroes to obtain the
same virtual resolution 
(this produces a change in the SNR and therefore a re-calibration 
of the spectra is needed).

%% \ref{fig:figure101}, \ref{fig:figure104} 
%%have the same resolution, $d \nu=8.1...\cdot 10^{-7}$,
%%that is $t_{obs}=14.1...$ days. This is the best we obtained with these
%%data, corresponding to the longest runs and also to very good data.

\begin{figure}
\centering\epsfig{file=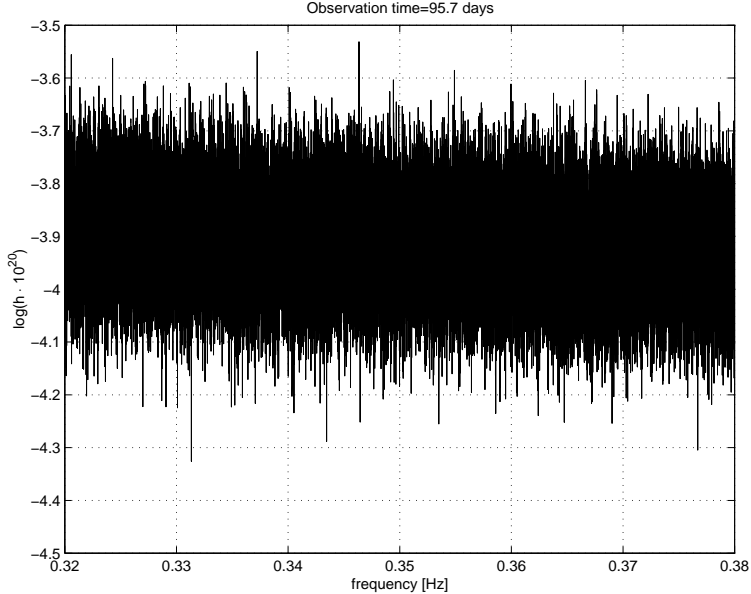,
width=10cm,clip=}
\caption{Average $\overline{h}$  from GC 
obtained using data from May to December
(95.7 days) in the
bandwidth 921.32-921.38 Hz.The x-axis is the frequency, in Hz-921 Hz} 
\label{fig:figure305z}
\end{figure}

The start times of these eleven runs, are days 
128.53, 137.29, 159.82, 171.43, 213.94, 225.32, 301.61, 312.37, 323.46, 339.68.
%%%%(and the virtual observation time 154 days).
%Fig. \ref{fig:figure305} shows the average 
%$\overline{h}$, from GC, in the 
%bandwidth 921.2-921.45 Hz and 
Fig.\ref{fig:figure305z} shows the average $\overline{h}$ for the 
%$\overline{h}$, from GC, in the 
frequency range 921.32-921.38.
In the frequency bandwidth 921.32-921.38 Hz over these 95.7 days 
the noise level is $(1.2 \pm 0.2) \cdot 10^{-24}$, well in
agreement with the expected value.
No lines with amplitude greater
than $\overline{h}=2.9 \cdot 10^{-24}$ are apparent. %%%$\nu=921.3467...$Hz. 

The standard deviation is a factor 2.5 lower than the standard deviation
obtained using only 21 days, thus we expect the efficiency of detection over
the 95 days be of the order of the unity 
even for signals of  $\overline{h}=2-3~10^{-24}$.

Thus, we exclude the possibility that, in the GC, 
there are sources 
having a spindown age $\tau \ge 10{^8}$ years,  
%questo dovrebbe essere 95 giorni incoerenti 
%ottenuti circa mediando 11 volte 7 gg
emitting signals with frequency 
in the range 921.32-921.38 Hz and strength (on Earth) greater 
than or equal to $\overline{h}=2.9~10^{-24}$.

%%Fig.\ref{fig:figure9_new_a} shows the $\overline{h}$ in the frequency range
%%921.2-921.45. 

%\par
%\begin{figure}
%\centering\epsfig{file=figure305.eps,
%width=10cm,clip=}
%\caption{Average $\overline{h}$  from GC 
% obtained using data from may to dicember 1991
%over 70.7...days, in the
%bandwidth 921.2-921.45 Hz. The figure is normalized to give the signal
%amplitude. Doppler correction from GC has been applied on the data.} 
%\label{fig:figure305}
%\end{figure}

%%\begin{figure}
%%\centering\epsfig{file=figure9_new_a.eps,
%%width=10cm,clip=}
%%\caption{Average $\overline{h}$  from GC 
%%obtained using data from May to December
%%(95.7 days) in the
%%bandwidth 921.2-921.45 Hz.The x-axis is the frequency, in Hz} 
%%\label{fig:figure9_new_a}
%%\end{figure}

%%eleven run, with average duration of seven days. porta a sigma di
%%0.3 e-24

\subsection{Point registration on the spectra}

The analyzed period consists of 51 runs, leading to 51 spectra of 
different resolution.
It is not convenient to average these spectra as done before for the
eleven longest ones, because now their duration are very different
one from each other.
They can be analyzed using various
methods- for example, by looking for patterns in the time evolution of their
spectral lines.
However this kind of analysis would require 
algorithms which are rather more involved than those used
in the present analysis (such algorithms are presently under 
investigation \cite{brady,map}).
The analysis here is also complicated by the fact that the
different spectra have different resolutions and thus 
different SNRs, for any given signal.

We restricted our search to a source in the GC emitting a signal
at constant frequency during the observation time.
We have tracked
all the local maxima in each spectrum obtained by setting a 
threshold\cite{map} at SNR $\simeq$ 4.
If a spectral line from the GC were present, it should show up  
in all the spectra (at various SNRs) at the same frequency.

\begin{figure}
\centering\epsfig{file=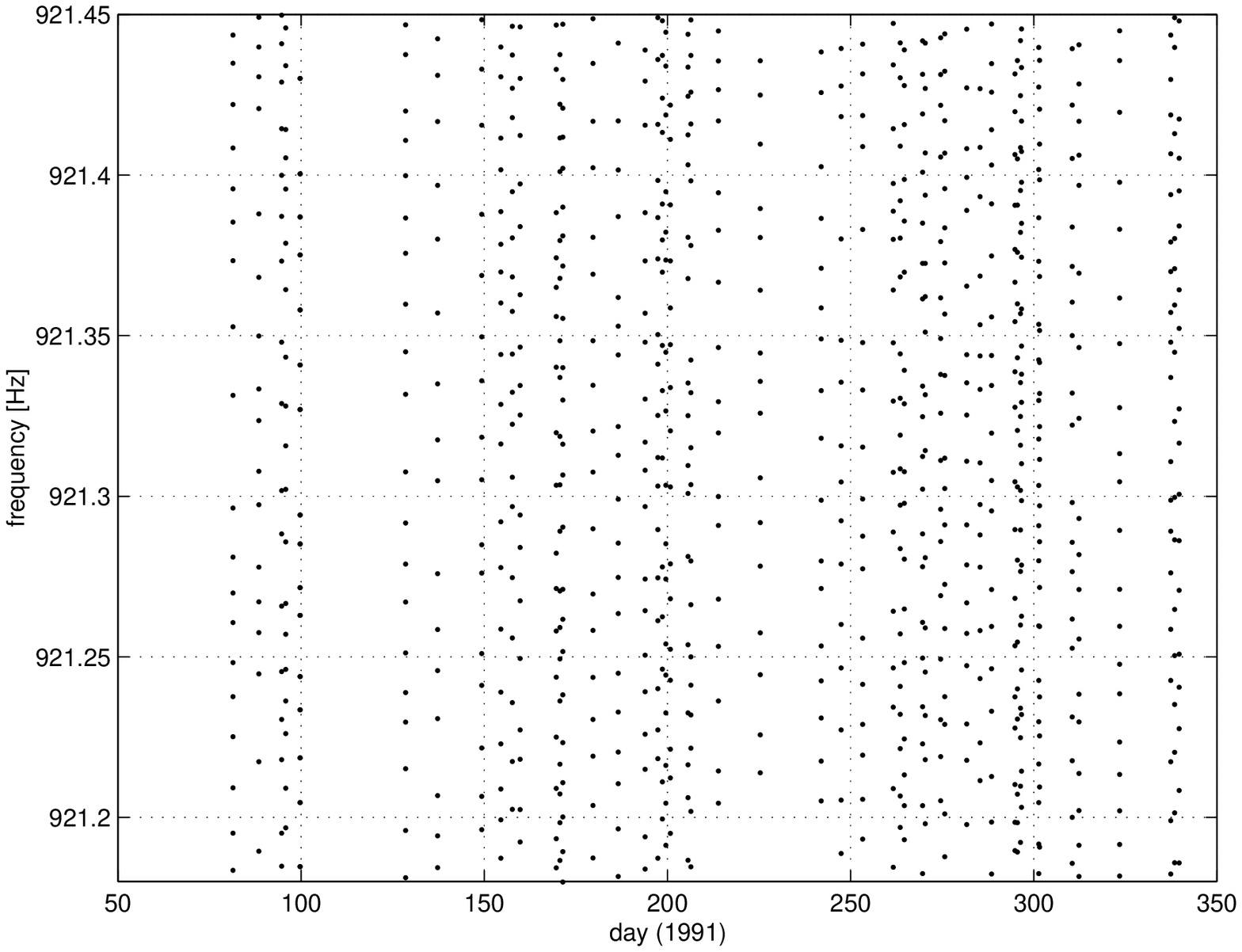,
width=12cm,clip=}
\centering\epsfig{file=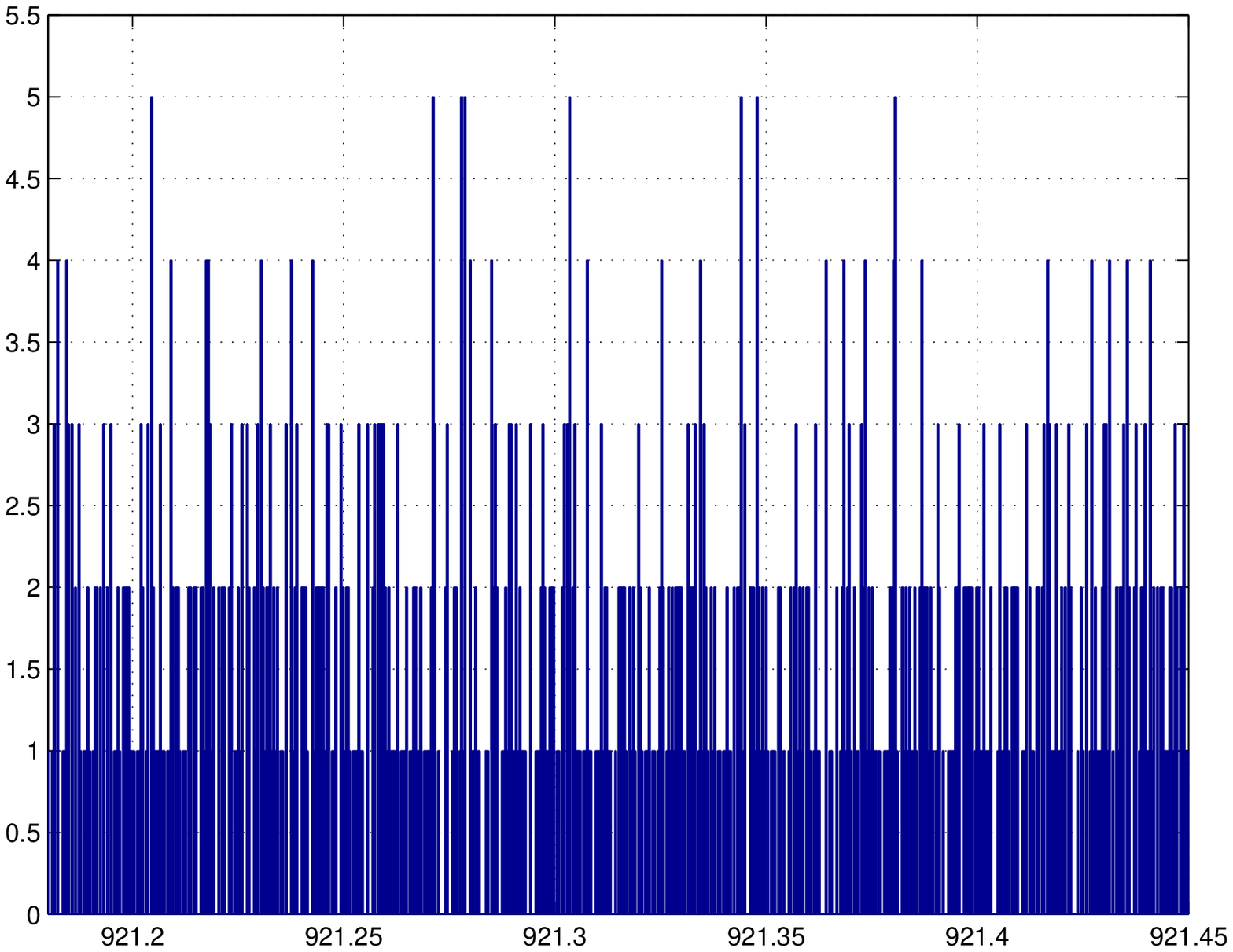,
width=12cm,clip=}
\caption{The incoherent analysis over 51 spectra from May to December.
The upper plot shows the time-frequency behaviour of the local maxima selected
in each spectrum. The lower plot is their histogram versus frequency.} 
\label{fig:point51}
\end{figure}

\begin{figure}
\centering\epsfig{file=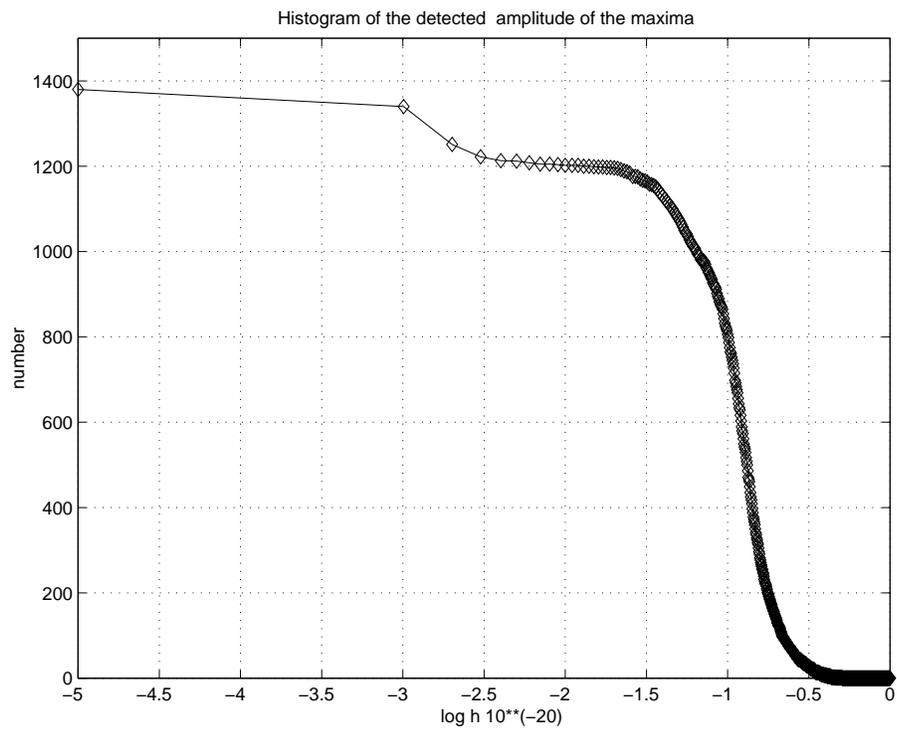,
width=12cm,clip=}
\caption{The figure shows the histogram of the (log) amplitudes of the 
selected local maxima in the 51 spectra} 
\label{fig:istoamp51}
\end{figure}

Fig. \ref{fig:point51} shows (top) the time-frequency plot of the selected
maxima and their histogram (bottom). 
The resulting histogram is flat and hence no evidence of straight 
horizontal lines is present in the data.
However, as shown in Fig.\ref{fig:istoamp51}, the sensitivity of this analysis
is much poorer than the previous method, as 
almost all these selected maxima have amplitude greater than $10^{-22}$.
%%Due to the different sensitivity 
%%for the various spectra the above null result
%%applies to lines visible with the spectra obtained on the shortest time
%%periods
%%%%togliere l' ultima frase, non e' cosi' 

\par
\newpage
\section{Conclusions}
\label{final}
A first analysis of the data obtained with the Explorer detector in 1991 was
performed, with the aim of searching for continuous g.w..
The analysis was limited to the frequency range 921.32-921.38 Hz, 
which contained the plus
resonance of the detector, where sensitivity was highest.
Doppler corrections on the g.w. frequency were made under the 
assumption that the source was still in the GC without 
any intrinsic frequency spin down.

No signals were observed with amplitude greater than $\overline{h}=
2.9~10^{-24}$, using data collected over 95.7 days, for a source located
at $\alpha=17.70 \pm 0.01$ hours and $\delta=-29.00 \pm 0.05$ degrees,
having a spindown parameter $\tau \ge 10{^8}$ years
(that is $\dot{p} \le 1.7 \cdot 10^{-19}$ s/s).  

The procedures adopted here can be applied to any assumed position in the 
sky of a g.w. source, for a greater frequency range, or even for a frequency
correction at the source due to spin down and Doppler effects.

\par
\newpage
%%\appendix
\appendix
\section*{\label{app:doppler} 
APPENDIX A:  Frequency resolution of the basic FFTs}

Here we want to show how we chose the spectral
resolution for our spectral data base \cite{livas,sergio}.

If $\overline{\nu}$ is the emission
frequency of the source then the observed frequency is given by
the (simplified) formula

\beq \nu=\overline{\nu} (1 \pm (v/c)) \label{doppler} \eeq
where $v=|\vec{v}_{rot}+\vec{v}_{orb}|$ is the relative 
velocity of the antenna and source,
due to $\vec{v}_{rot}$, daily rotation of the Earth, 
and to $\vec{v}_{orb}$, orbital
motion around the Sun.

{\bf Orbital~motion,~annual:}
\beq \vec{v}_{orb}=|v_{orb}| \cdot \cos(\Omega_{orb} t)\eeq
\beq |v_{orb}|=\Omega_{orb} R_0=2 \pi R_0/T_{orb}= 2 \pi R_0/(365 \cdot 86400)=
29.772~~km/s\eeq 
$R_0=149504201~km$;
$\Omega_{orb}=2 \pi/(86400 \cdot 365)=1.99 \cdot 10^{-7} rad/s$;
$|v_{orb}|/c \simeq 0.9933 \cdot 10^{-4}$.

\beq
{d \nu_{orb} \over{dt}}=\overline{\nu}~(1/c)~\cos\beta_{ecl}~
{d \vec{v}_{orb} \over{dt}}
\eeq
\beq
{d \vec{v}_{orb} \over{dt}}=R_0 \Omega^2_{orb}\cdot \cos(\Omega_{orb} t)
=\vec{v}_{orb} \Omega_{orb}
\eeq

\beq
{d \nu_{orb} \over{dt}}|_{max}=\overline{\nu}~(|v_{orb}|/c)~\cos\beta_{ecl}~
\Omega_{orb}=
\overline{\nu}~(|v_{orb}|/c)~\cos\beta_{ecl}~ {2 \pi \over{T_{orb}}}~~{Hz \over {s}}
\label{almassimo}\eeq

\beq \simeq \overline{\nu} \cdot 1.9778 \cdot 10^{-11}~\cos\beta_{ecl}~~{Hz \over {s}}
\label{almassimo1}\eeq

\indent 

{\bf Daily~rotation:}
\beq \vec{v}_{rot}=|v_{rot}| \cdot \cos(\Omega_{rot} t)\eeq
\beq |v_{rot}|=\Omega_{rot} R_T=2 \pi R_T/T_{rot}=
2 \pi R_T/86400=0.4636~km/s\eeq
$R_T=6378.388~km$;
$\Omega_{rot}=2 \pi/86400=7.2 \cdot10^{-5}~rad/s$;
$|v_{rot}|/c \simeq 1.5467 \cdot 10^{-6}$.

\beq
{d \nu_{rot} \over{dt}}|_{max}=\overline{\nu}~(|v_{rot}|/c)~\cos\beta_{ter}
~\Omega_{rot}=
\overline{\nu}~(|v_{rot}|/c)~\cos\beta_{ter}~ {2 \pi \over{T_{rot}}}~~{Hz \over {s}}\eeq

$$\simeq \overline{\nu} \cdot 11.244 \cdot 10^{-11}~\cos\beta_{ter}~~{Hz \over {s}}$$

We will use here the symbols $|\Delta \nu_{max}|_{rot}$ and
e $|\Delta \nu_{max}|_{orb}$ 
for the maximum frequency shift, in the time interval $\Delta T$.

\beq {|\Delta \nu_{max}|_{rot} \over {(\overline{\nu}~\Delta T)}}=11.244 \cdot 10^{-11}
\cos\beta_{ter}~~s^{-1} \label{rotazione}\eeq
if $\Delta T$ less than 1 day.

\beq {|\Delta \nu_{max}|_{orb} \over {(\overline{\nu}~\Delta T)}}=1.977 \cdot 10^{-11}
\cos\beta_{ecl}~~s^{-1} \label{rivoluzione}\eeq
if $\Delta T$ less than 1 year.

%$${[Hz/s]_{rot} \over{[Hz/s]_{orb}}}=5.685~~~~(T < 1~giorno),$$
%cioe' l' effetto della rotazione e' piu' forte di quello di rivoluzione di
%circa un fattore 5.68.

\beq
(|\Delta \nu_{max} / \Delta T|_{rot} +  |\Delta \nu_{max} / \Delta T|_{orb})
\cdot T ~~~~(T < 1~day)\label{effetto_tot}\eeq

 \beq  
(|\Delta \nu_{max} / \Delta T|_{rot} +  |\Delta \nu_{max} / \Delta T|_{orb})
\cdot T < \delta_{\nu_f} \eeq 

$$1/T > T \cdot (a+b)$$

where

\indent
$a= \overline{\nu} \cdot 11.244 \cdot 10^{-11} \cos \beta_{ter}$ ${Hz \over {s}}$, 
(daily rotation)

\indent
$b= \overline{\nu} \cdot 1.977 \cdot 10^{-11} \cos \beta_{ecl}$ ${Hz \over {s}}$, 
(annual orbital motion)

%%$$(a+b)= \overline{\nu} \cdot 13.22 \cdot 10^{-11} \cos \beta_{ecl} ~~~
%%{Hz \over {s}}$$

$$T^2 <1/(a+b)$$
%%$$T < 8.6972 \cdot 10^{4}/\sqrt{\overline{\nu}~\cos \beta_{ecl}}~~~s $$
$$T < 8.6972 \cdot 10^{4}/\sqrt{\overline{\nu}}~~~s $$

If we consider $\overline{\nu}=1000~Hz$

%%\beq T < {2749.9 \over{\sqrt{\cos \beta_{ecl}}}} ~~s \label{tscelto}\eeq
\beq T < {2749.9} ~~s \label{tscelto}\eeq

In our data, the sampling time is $\delta_{t_f}=18.176~ms$ then, using
$T=2382.35~s$, that is 0.6617 hours, 
we have a frequency resolution 
$$\delta_{\nu_f} =0.41975~mHz$$

that is $2^{17}$ = 131072 samples in each periodogram.

\appendix
\section*{\label{app:clock} APPENDIX B:  Timing precision}

We note here that precision in both the absolute time and the sampling 
frequency during the observation time $t_{obs}$ is 
crucial when combining the basic FFTs to construct one very long spectrum.

If $\nu$ is the frequency of the wave, $\tau={1 \over{\nu}}$ the wave period,
$\epsilon_{\nu}={\Delta \nu \over{\nu}}$ the relative frequency error
($\Delta \nu$ represents here the wave frequency indetermination due to the 
acquisition and analysis procedures)
then $$\Delta \tau=\epsilon_{\nu} \cdot \tau$$

After time $t_m$ we have a phase error of
$${\Phi \over{2 \pi}} =t_m \Delta \nu = t_m \nu \epsilon_{\nu} $$

Hence, if we want a phase error ${\Phi \over{2 \pi}} < \simeq 0.01$ in
$t_m=1~year=3.1 \cdot 10^{7}~s$ 
we need $\epsilon_{\nu} {< \simeq 3 \cdot 10^{-10} {\over{\nu}}}$

Hence, if $\nu = 1~kHz$:
$$\epsilon_{\nu} < \simeq 3 \cdot 10^{-13}$$
$$\Delta \tau < \simeq 10^{-5} s/year$$

\appendix
\section*{\label{app:adva} 
APPENDIX C:  Practical issues in the construction of the data base}

\begin{itemize}

\item
Each FFT is computed using $2N$ data, sampled with sampling time $\Delta
t$. The data are windowed, in the time domain, before the Fourier
transform. This means that the data $y_i$ are
multiplied by $w_i=A-B \cos{(i)} + C \cos{(2 i)}$,
 where $i =(0,2N-1)\cdot 2\pi/(2N-1)$.
In the present analysis we have used 
a Hamming window, that is $A=0.54, ~B=0.46$ and $C=0$. 
%%%%%{ \it devo spiegare perche' Hamming, o comunque rimandare ad un testo}

\item 
The FFTs are  stored in units of  strain/ $\sqrt {\mbox{Hz}}$, and are
normalized so that their squared modulus is the spectrum. 

\item 
The basic FFTs of the data base overlap for half their length.  The
time duration of each FFT is $t_0= 2N \Delta t$, and a new FFT is done
after time $t_0/2$. This is important since it avoids distortions in
the final time domain sequence - this is the well known "overlap-add"
method, described in many data analysis textbooks. For the Explorer
detector we thus have 110 FFTs over 36 hours with $\delta \nu=0.41...$
mHz.

\item
We select the frequency bandwidth to be analyzed and we add zeroes
to construct the analytical signal.
These data should be to a power of two, to allow use of a Fast Fourier
algorithm.
The chosen bandwidth should be wide enough to include all the frequencies
we expect to observe due to the Doppler shift from the
given source, during the time of
observation.

\item
After the bandwidth has been selected, the data (still in the frequency domain)
should be windowed, to avoid edge effects in the transformed data.

\item
The selected data are then transformed to return to the time
domain. At this stage we must remove the window used in the data
when constructing the FFT data base,
 by simply dividing the new time domain data by $w(t)$.
This operation recovers the original time data (subsampled) because the
only regions where the division might not work are the edges of the data
stream, where the value of the function 
$w(t)$ may be zero, depending on the kind of
window used (a problem which, of course, has been
overcome by the overlapping of the FFTs).

\item
If an FFT under consideration is vetoed or if it is
missing, then the data are set to zero.

\item
Each new group of
time domain data is appended to the previous groups, after elimination of
the overlapped data. Since the overlapping concerns half the data
we eliminate 1/4 of the data at the beginning and end of each stream.
The first 1/4 data in the first FFT and the last 1/4 in the
last FFT can be discarded.
The data of missing or vetoed periods, set to zero as
explained above, are appended to the others in the same way.

\item
At this stage we have a subsampled time domain data stream, which represents
the analytical signal associated with the original data. 

\item
Now we can take into account the Doppler shift and correct the data
as previously explained.

\item
The final step is to calculate the power spectrum from this
subsampled time domain data (after data windowing in the time
domain)

\end{itemize}

\appendix
\section*{\label{app:para} APPENDIX D:  Uncertainty in the source parameters}

\subsubsection*{Uncertainty in the source position parameters}

\begin{figure}
\centering\epsfig{file=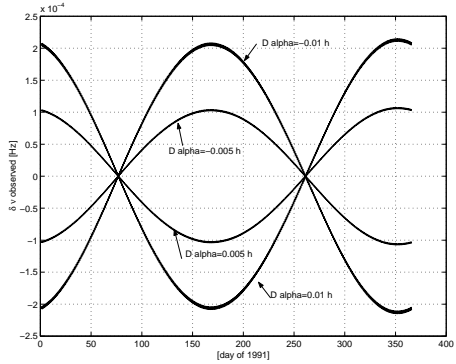,
width=6.0cm,clip=}
\caption{The graph shows the difference in the observed frequency on Earth
between a signal in the GC and signals coming from
nearby coordinates. The x-axis are days of 1991.
The y-axis are the frequencies, from  $-2.5 \cdot 10^{-4}$
to $2.5 \cdot 10^{-4}$ Hz}
\label{fig:dalpha}
\end{figure}

In this analysis we used 
 the coordinates $\alpha=17.7$ hour,$\delta=-29.0$ deg to define the GC.
In order to calculate the region of the sky effectively covered by this 
definition it was necessary to study the effect on  the analysis 
of a source not being ``exactly'' in the GC, since the
 frequency modulation depends on the precise location of the source.
To get an idea of the problem in 1991 we plotted (Fig. \ref{fig:dalpha}) 
the difference in the observed frequencies on Earth
between a signal from the GC 
and signals coming from sources at nearby coordinates. 

From the graph
it is easy to see that a difference in the right ascension of $\pm 0.01$ hours
leads to a maximum difference in the observed frequency of 
$\simeq 2 \cdot 10^{-4}$ Hz.
This mismatch is at a maximum twice a year, at the beginning of June and at the
beginning of December.
Thus, we studied the effect of the mismatch 
during a run 
%%%%%%of 8 days
 in December, when it was at a maximum.

\begin{figure}
\centering\epsfig{file=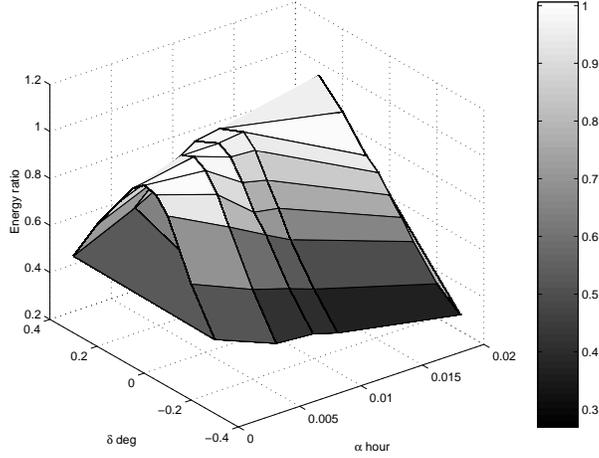, width=8cm,clip=}
\caption{3-dimensional plot of the data in Table I. 
The z-axis is the energy of the signal integrated over the three bins 
(maximum,previous, next). The x and y-axes are the mismatch in right 
ascension (hours) and in declination (degrees)} 
\label{fig:curva3}
\end{figure}

Table I shows the results for uncertainties both in right ascension
and in declination.
The first column gives the error in right ascension
($\alpha$ in hours); 
the second, the error in declination ($\delta$ in degrees); the third, the
energy of the spectrum of the signal, $\mbox{1/Hz}$, in the frequency bin 
of its maximum and in the previous and next bins nearest to the maximum; 
the fourth column gives the difference in the frequency of the signal
compared to the nominal, expressed in number of bins
(one bin is $8.1 \cdot 10^{-7}$ Hz).
Fig.\ref{fig:curva3} is the corresponding 3-dimensional plot. The z-axis
is the energy of the signal, integrated over the three bins.
%%%%%%%%%%The top lines in the figure are the points corresponding to z-axis=1.

%\begin{figure}
%\centering\epsfig{file=curva1.eps, width=8cm,clip=}
%\caption{3-dimensional plots of the data in Table 1. 
%The z-axis is the energy of the signal integrated over the three bins 
%(maximum,previous, next). The x and y-axes are the mismatching in right 
%ascension (hours) and declination (degrees).
%The top lines in the figure are the points corresponding to z-axis=1. } 
%\label{fig:curva1}
%\end{figure}

\par
\newpage

\begin{table}
\caption{Results for uncertainties in right ascension
and in declination}
\begin{center}
\begin{tabular}{lcccccc}
\multicolumn{6}{c}{\bf{Table I}} \\[0.5 cm] \hline
$\Delta \alpha~h$,  & \ & $\Delta \delta~deg$, 
 & \ &  signal energy & \ &  $\Delta f$ (n bin)       \\
 0.000 & \ &  -0.30 & \ &  0.25;    0.31;     0.28 & \ &   +13 \\
 0.000 & \ &  -0.20 & \ &  0.24;    0.46;     0.43 & \ &    +8 \\
 0.000 & \ &  -0.10 & \ &  0.35;    0.79;     0.37 & \ &    +4 \\
 0.000 & \ &  -0.05 & \ &  0.40;    0.95;     0.26 & \ &    +2 \\
0.000 & \ &   0.000 & \ &  0.46;    1.00;     0.17 & \ &    +0 \\
0.000 & \ &   +0.05 & \ &  0.51;    0.92;     0.13 & \ &    -2 \\
0.000 & \ &   +0.10 & \ &  0.56;    0.74;     0.13 & \ &    -4 \\
0.000 & \ &   +0.20 & \ &  0.31;    0.51;     0.35 & \ &    -9 \\
0.000 & \ &   +0.30 & \ &  0.24;    0.33;     0.29 & \ &    -14 \\
0.005 & \ &   -0.30 & \ &  0.26;    0.26;     0.15 & \ &   +143 \\
0.005 & \ &   -0.20 & \ &  0.35;    0.37;     0.23 & \ &   +139 \\
0.005 & \ &   -0.10 & \ &  0.31;    0.59;     0.39 & \ &   +134 \\
0.005 & \ &   -0.05 & \ &  0.40;    0.72;     0.35 & \ &   +131 \\
0.005 & \ &   0.000 & \ &  0.47;    0.85;     0.27 & \ &   +129 \\
0.005 & \ &   +0.05 & \ &  0.50;    0.95;     0.19 & \ &   +128 \\
0.005 & \ &   +0.10 & \ &  0.50;    0.96;     0.14 & \ &   +126 \\
0.005 & \ &   +0.20 & \ &  0.51;    0.66;     0.18 & \ &   +127 \\
0.005 & \ &   +0.30 & \ &  0.29;    0.43;     0.33 & \ &   +116 \\
0.008 & \ &  -0.30  & \ &  0.21;    0.26;     0.18 & \ &   +220 \\
0.008 & \ &  -0.20  & \ &  0.24;    0.34;     0.29 & \ &   +206 \\
0.008 & \ &  -0.10  & \ &  0.43;    0.47;     0.25 & \ &   +202 \\
0.008 & \ &   -0.05 & \ &  0.19;    0.55;     0.52 & \ &   +199 \\
0.008 & \ &  0.000  & \ &  0.18;    0.71;     0.53 & \ &   +197 \\
0.008 & \ &  +0.05  & \ &  0.15;    0.88;     0.52 & \ &   +195 \\
0.008 & \ &  +0.10  & \ &  0.13;    0.96;     0.51 & \ &   +193 \\
0.008 & \ &  +0.20  & \ &  0.23;    0.79;     0.50 & \ &   +189 \\
0.008 & \ &  +0.30  & \ &  0.34;    0.50;     0.35 & \ &   +184 \\
0.010 & \ &  -0.30  & \ &  0.16;    0.23;     0.21 & \ &   +271 \\
0.010 & \ &   -0.20 & \ &  0.26;    0.31;     0.24 & \ &   +266 \\
0.010 & \ &   -0.10 & \ &  0.24;    0.44;     0.38 & \ &  +261 \\
0.010 & \ &   -0.05 & \ &  0.29;    0.56;     0.38 & \ &  +260 \\
0.010 & \ &   0.000 & \ &  0.35;    0.67;     0.36 & \ &  +255 \\
0.010 & \ &   +0.05 & \ &  0.42;    0.79;     0.33 & \ &  +256 \\
0.010 & \ &   +0.10 & \ &  0.43;    0.92;     0.27 & \ &  +254 \\
0.010 & \ &   +0.20 & \ &  0.36;    0.95;     0.24 & \ &  +250 \\
0.010 & \ &   +0.30 & \ &  0.40;    0.59;     0.29 & \ &  +246 \\
0.020 & \ &   -0.30 & \ &  0.11;    0.18;     0.15 & \ &  +429 \\
0.020 & \ &   -0.20 & \ &  0.21;    0.23;     0.12 & \ &  +426 \\
0.020 & \ &   -0.10 & \ &  0.23;    0.30;     0.23 & \ &  +421 \\
0.020 & \ &   -0.05 & \ &  0.29;    0.34;     0.25 & \ &  +419 \\
0.020 & \ &   0.000 & \ &  0.35;    0.40;     0.24 & \ &  +417 \\
0.020 & \ &   +0.05 & \ &  0.43;    0.45;     0.22 & \ &  +415 \\
0.020 & \ &   +0.10 & \ &  0.16;    0.52;     0.49 & \ &  +412 \\
0.020 & \ &   +0.20 & \ &  0.08;    0.68;     0.62 & \ &  +408 \\
0.020 & \ &   +0.30 & \ &  0.64;    0.78;     0.12 & \ &  +405 \\

%%%\hline
\end{tabular}
\end{center}
\end{table}
\vspace{0.5cm}

It is not easy to arrive at a  general conclusion, because 
the final effect depends on the
mismatch on both the parameters and, in some cases, when the two parameters
act in opposite direction, the final result is better compared to a
mismatch in only one of the two parameters. This is why, for example,
the result when $\Delta \alpha$=0.01 hour and  $\Delta \delta$=0.3 deg
is better than
$\Delta \alpha$=0.01 hour and  $\Delta \delta$=0 deg.
Anyway, assuming the analysis is valid even when there is a worsening
by a factor of 2 in SNR we may conclude
that the region of the sky under study is definitely within 
either the volume
$\alpha=17.7 \pm 0.01$ hours and $\delta=-29.0 \pm 0.05$ degrees
(0.01 hours=0.15 degrees), or the volume
$\alpha=17.7 \pm 0.005$ hours and $\delta=-29.0 \pm 0.2$ degrees
We also note that the observed frequency shift of the 
maximum is consistent with that which would be 
expected (compare, for example, the results
in Table I with those in Fig. \ref{fig:dalpha}).

\par
\newpage
\subsubsection*{Uncertainty in the source frequency}

\begin{figure}
\centering\epsfig{file=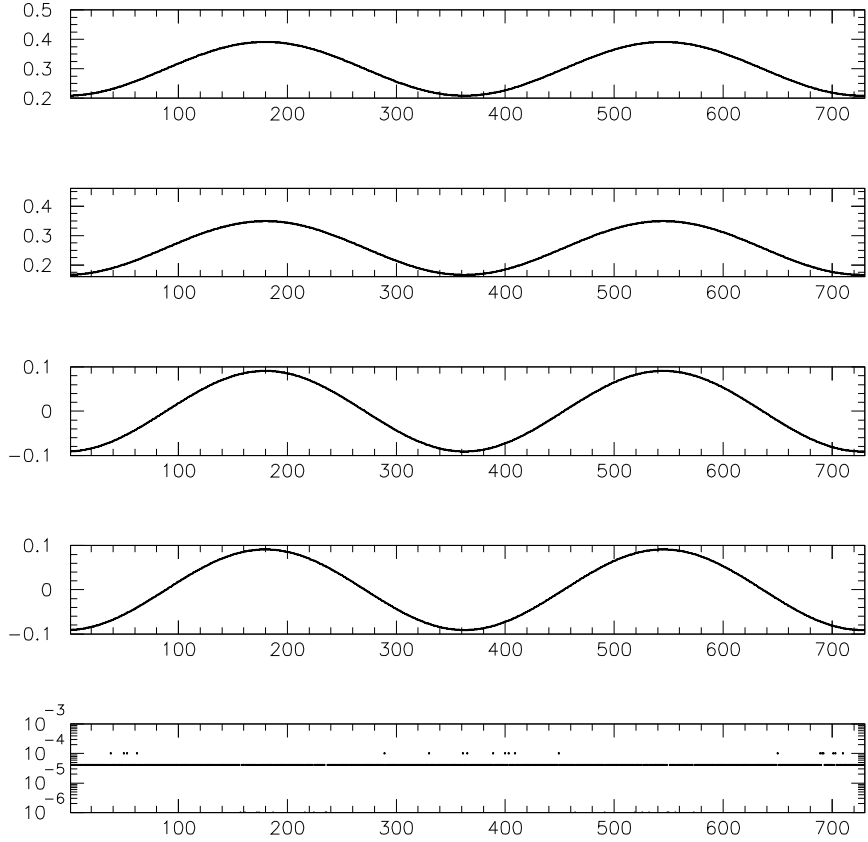,
width=12cm,clip=}
\caption{The top figure shows the observed $\nu_{1d}$ frequency-921 Hz, 
assuming a source in the GC emitting at 
$\nu_1$=921.300 Hz. The x-axis is in days. 
The second figure shows the same
for a signal at $\nu_2$=921.257 Hz. The third and the forth
are  $\nu_1-\nu_{1d}$ and $\nu_2-\nu_{2d}$. The bottom figure  is the
difference between the third and the fourth figures.} 
\label{fig:sin_test}
\end{figure}

To test the extent to which the analysis depends 
on knowledge of the intrinsic
frequency of the source, we did a simulation of a 
spectrum of 14.1 days, by introducing a signal at 921.3 Hz and correcting it,
during Doppler removal, using 921.2 and 921.4 Hz (that is with an
error of $\pm 0.1$ Hz). 
From Table II it is easy to see that 
there are no significant differences in the resulting spectra. Thus the
final result is only slightly affected by even a
very ``big'' error such as this.
This finding is important because it allows us to analyze just a 
set of discrete frequencies, for example just 1/100  of the frequencies
in the original FFTs ($0.419...$ mHz). 
\begin{table}
\caption{Results for uncertainty in the source frequency.
The first column gives the error in the frequency
($\Delta \nu$ in Hz); 
; the second, the
energy of the spectrum of the signal, $\mbox{1/Hz}$, in the frequency bin 
of its maximum and in the previous and next bins nearest to the maximum; 
the third column gives the difference in the frequency of the signal
compared to the nominal, expressed in number of bins
(one bin is $8.1 \cdot 10^{-7}$ Hz)}
\begin{center}
\begin{tabular}{lccccc}
\multicolumn{5}{c}{\bf{Table II}} \\[0.5 cm] \hline
$\Delta \nu$ & \ &   signal energy & \ & $\Delta f$ (n bin)       \\
   0.0 & \ &  (0.46) 1.00 (0.17) & \ &     +0 \\
  +0.1 & \ &  (0.29) 0.92 (0.39) & \ &     +1  \\
  -0.1 & \ &  (0.71) 0.73 (0.15) & \ &     -1  \\ 

%%%%\hline
\end{tabular}
\end{center}
\end{table}
\vspace{0.5cm}
It can be shown that this property is intrinsic to the nature of the
Doppler correcting factor (eq. \ref{correctingfactor}), which depends 
on the
{\it difference} between the intrinsic and the observed frequency.
Fig.\ref{fig:sin_test} shows, for example, the difference (over two years)
between the terms $\nu_1-\nu_{1d}$ and $\nu_2-\nu_{2d}$, where
$\nu_1=921.300$ Hz, $\nu_2=921.257$ Hz, $\nu_{1d}$ and $\nu_{2d}$ are the
observed frequencies at the detector, assuming the source to be in the GC.
This difference is of the order of $3 \cdot 10^{-5}$.

\par
\newpage
%%%%%%%%%%%\section*{References}

\end{document}